\documentclass[11pt]{article}
\usepackage{amssymb,amsmath}
\usepackage{bm} 
\usepackage{booktabs} 
\usepackage{array}
\usepackage{latexsym}
\usepackage{graphicx}
\usepackage{color}
\usepackage{datetime}
\usepackage[nosort]{cite}
\usepackage{verbatim}
\usepackage{enumerate}
\usepackage{chngpage} 
\usepackage{mathrsfs}
\usepackage{euscript}
\usepackage{psfrag}

\usepackage{mciteplus}

\usepackage[colorlinks=true,      linkcolor=blue,      urlcolor=blue,
            filecolor=blue,      citecolor=blue,       pdfstartview=FitH,
                        pdfpagemode=UseNone,      bookmarksopen=true]{hyperref}  
\usepackage[all]{hypcap}     

\usepackage{tensor}
\usepackage{physics}
\usepackage{tikz}

\newcommand\snowmass{\begin{center}\rule[-0.2in]{\hsize}{0.01in}\\\rule{\hsize}{0.01in}\\
\vskip 0.1in Submitted to the  Proceedings of the US Community Study\\ 
on the Future of Particle Physics (Snowmass 2021)\\ 
\rule{\hsize}{0.01in}\\\rule[+0.2in]{\hsize}{0.01in} \end{center}}

\renewcommand{\comm}[1]{} 

\def\({\left(}
\def\){\right)}
\def\[{\left[}
\def\]{\right]}

\def\One{{\hbox{ 1\kern-.8mm l}}}

\def\barray{\begin{array}}
\def\earray{\end{array}}
\def\be{\begin{equation}}
\def\ee{\end{equation}}
\def\bea{\begin{eqnarray}}
\def\eea{\end{eqnarray}}
\def\bal{\begin{align}}
\def\eal{\end{align}}

\numberwithin{equation}{section} 

\definecolor{cardinal}{rgb}{0.6,0,0}
\definecolor{darkgreen}{rgb}{0,0.4,0}
\definecolor{golden}{rgb}{0.92, 0.7, 0}
\definecolor{midnight}{rgb}{0, 0, 0.5}
\definecolor{darkblue}{rgb}{0, 0, 0.7}
\definecolor{darkred}{rgb}{0.6, 0, 0}
\definecolor{purple}{rgb}{0.5, 0, 0.5}

\def\nicebox#1{\bigskip \framebox{\parbox{5.8 in}{{\it #1}}}\bigskip }

\begin{document}

\phantom{AAA}
\vspace{-15mm}

%
%

\vspace{1.9cm}

\begin{center}

{\huge {\bf  Snowmass White Paper: }}\\
{\huge {\bf \vspace*{.25cm} Micro- and Macro-Structure of Black Holes }}\\

\vspace{0.5cm}

{\large{\bf {Iosif Bena$^1$,~Emil J. Martinec$^{2}$ ,~Samir D. Mathur$^{3}$  and  Nicholas P. Warner$^{1,4,5}$}}}

\vspace{0.5cm}

$^1$Institut de Physique Th\'eorique, \\Universit\'e Paris-Saclay, CNRS, CEA,
91191, Gif-sur-Yvette, France \\[8 pt]

$^2$Enrico Fermi Institute\ and Department\ of Physics, \\
University of Chicago,  5640 S. Ellis Ave.,
Chicago, IL 60637-1433, USA\\[8 pt]

$^3$Department of Physics,\\ 
The Ohio State University,\  Columbus,  OH 43210, USA \\[8 pt]

$^4$Department of Physics and Astronomy, 
and $^5$Department of Mathematics \\
University of Southern California,  Los Angeles, CA 90089-0484, USA

\vspace{0.6cm} 
{
\small
\upshape\ttfamily 
iosif.bena @ ipht.fr~,~
e-martinec @ uchicago.edu~,\\
mathur.16 @ osu.edu~,~
warner @ usc.edu} \\

\vspace{0.6cm}
 
{\textsc{Abstract}}\\
\end{center}
The black-hole information paradox provides a stringent test of would-be theories of quantum gravity.   String theory has made significant progress toward a resolution of this paradox, and has led to the fuzzball and microstate geometry programs. The central thesis of these programs is that only string theory has sufficiently many degrees of freedom to resolve black-hole microstructure, and that horizons and singularities are artifacts of attempting to describe gravity using a theory that has too few degrees of freedom to resolve the physics.    Fuzzballs and  microstate geometries recast black holes within string theory as horizonless and singularity-free objects that not only resolve the paradox but provide new insight into the   underlying microstructure.  We give an overview of this approach, summarize its current status and describe future prospects and insights that are now within reach.

\snowmass

\begin{adjustwidth}{3mm}{3mm} 
 
\vspace{-1.4mm}
\noindent
%
%

%
\end{adjustwidth}

\thispagestyle{empty}
\newpage


\baselineskip=17pt
\parskip=5pt

\setcounter{tocdepth}{2}
\tableofcontents

\baselineskip=15pt
\parskip=3pt


\vskip 1cm
\hrule
\vskip 2cm

\setcounter{page}{1}
\section{Overview}
\label{sec:Overview}

LIGO has now detected over 50 mergers of black holes, and the Event Horizon Telescope is producing images of accretion disks with higher and higher resolution.    These observations are utterly remarkable as direct measurements of black holes; furthermore, they represent a phenomenal validation of General Relativity (GR).  The wave-forms measured by LIGO show that GR  captures the highly non-linear extremes of black-hole mergers with stunning fidelity.  It is therefore very natural to ask, why, at such a pinnacle of success, should we be seeking to replace GR with a ``better'' theory of gravity.

Black holes are defined to be objects whose gravity is strong enough to trap light. In GR, a black hole not only traps light but also traps all other matter and information inside a surface of no return~-- the event horizon. As a result, the exterior of a black hole (outside the horizon) is independent of how, and from what, the black hole formed.  Thus, the exterior structure  is completely determined by its long-range parameters, like mass, charge and angular momentum.  This black hole uniqueness is directly related to the fact that, in GR,  the horizon region is a vacuum: Anything near the horizon will be swept into the center of the black-hole in the time it takes light to cross the scale of the horizon ($\sim 10^{-5}$ seconds for a solar-mass black hole).

In 1975, Hawking showed that the correct  description (in GR) of the quantum vacuum around an event horizon leads to the emission of Hawking Radiation as a form of vacuum polarization \cite{Hawking:1975vcx}. Because this radiation originates from just above the horizon, the uniqueness of black holes in GR implies that Hawking radiation is universal, thermal and (almost) featureless. In particular, it is independent of how the black hole formed. Semi-classical back-reaction of this Hawking radiation also implies that the black hole will evaporate, albeit extremely slowly. 

This leads to the Information Paradox  \cite{Hawking:1976ra,Hawking:1982dj}:  The impossibility of reconstructing the interior state of a black hole (apart from mass, charge and angular momentum) from the exterior data, and thus from the final state of the Hawking radiation.  As a result, the evaporation process cannot be represented as a unitary transformation of states in a Hilbert space.  Hence black-hole evaporation, as predicted by GR and quantum field theory, is inconsistent with the foundational postulate of unitarity in quantum mechanics.  Based on its horizon area, the black hole at the core of the Milky Way should have about $e^{10^{90}}$ microstates.  From the outside, black-hole uniqueness implies that its state is unique, as would be the state of its Hawking radiation after the black hole has evaporated. The problem is therefore vast: $e^{10^{90}} \ne 1$.

In the 47 years since Hawking's original paper, there have been many attempts at resolving the information problem.  Working purely within quantum mechanics and GR has proven to be a dead end, except to reformulate the problem and highlight its extreme difficulty.  The paradox is also not the result of some unexpected failure of semi-classical physics.  Quite the opposite; the problem is at its sharpest for large black holes.   The horizon size is proportional to the mass, $m$, of the black hole, while gravitational tidal effects at the horizon fall off as $m^{-2}$. The conventional wisdom therefore suggests that space-time in the horizon region of a large, astrophysical black hole behaves like any other (nearly) flat piece of space-time.  Moreover, any matter near the horizon will be swept inside the black hole within the light-crossing time ($\sim m$) of the black hole, leaving behind a vacuum.  There are many theorems, like black-hole uniqueness, that show that it is ``impossible" to support any conventional matter, or other structure, near the horizon of a black hole.  Therefore, within the framework of GR and quantum mechanics, the near-horizon region will behave just like a normal piece of (nearly) flat space-time quantum vacuum. However, it is precisely this vacuum and the normalcy of the space-time at the horizon that leads to Hawking radiation and thus to the information problem.

In 2009, one of the authors of this white paper used quantum information theory to show that Hawking's information paradox could not be solved incrementally~-- It will require radical modification of the conventional wisdom \cite{Mathur:2009hf}. Prior to this, it had been a common belief that, because the evaporation time of a black hole is staggeringly large, the information about how the black hole formed could leak out very slowly via tiny corrections to GR (or to quantum field theory).  The ``small corrections theorem" \cite{Mathur:2009hf}, and its generalization in the  ``effective small corrections theorem'' \cite{Guo:2021blh}, are based on widely-accepted ideas of how quantum field theory is formulated near black holes.  These results showed that resolving the paradox requires ``order~$1$'' changes to horizon-scale physics.  Radical reinvention has now become an imperative.

Specifically, these theorems  tell us that to resolve the information problem one must give up (at least) one of three things:
\begin{enumerate}[\it (i)]
\setlength\itemsep{-0.1em}
\item Smooth horizons for black holes 
\item Unitarity of quantum mechanics in the presence of black holes
\item Locality of quantum mechanics in the presence of black holes 
\end{enumerate}

Many approaches have tried to evade the consequences of this theorem but fall afoul of it, ultimately (and usually implicitly) violating (ii) or (iii) in a vain attempt to preserve (i).   However, it is extremely hard to reconcile violations of (ii) or (iii) with quantum field theory and string theory.  On the other hand, extensive analysis and calculations over the last 25 years indicate that, within string theory, one can violate option (i) and still obtain consistent black-hole-like behavior from suitably constructed compact objects.  Indeed, the fuzzball paradigm and microstate geometry program seek a gravitational and quantum description of horizon-scale microstructure in terms of horizonless objects in string theory. This approach seems to be the most promising route to solving the information paradox because it addresses the detailed structure of the microstates, while including complete, general-relativistic, (quantum-) gravitational effects on the microstructure.  Conversely, these frameworks suggest that black holes, and the extreme blue-shifts of probes they generate, might provide a way to test string theory. 

Our purpose here is to give a broad overview of the physical ideas underpinning fuzzballs  and microstate geometries, and describe some of the highlights of the developments of these programs.  We are also preparing a somewhat longer review of these concepts for those who would like a little more detail \cite{BMMW}.

\section{Fuzzballs and microstate geometries: A broad perspective}
\label{Sect:perspective}

\subsection{Some background}
\label{ss:background}

The first real advance on the Information Paradox came with the work of Sen \cite{Sen:1995in} and Strominger and Vafa \cite{Strominger:1996sh}, who showed that a string theoretic description of a black hole has the capacity to count its microstates.  This was demonstrated in a very specialized limit: The black holes were supersymmetric, and thus had vanishing Hawking temperature, and the microstate counting was done in a limit in which the string coupling, and thus Newton's constant, $G_N \sim g_s^2$, vanishes.  The triumph was to show that the index states of the string system have an entropy that exactly matches the horizon area of the black hole at finite $G_N$.  Thus string theory seemes to have the fidelity needed to resolve black-hole microstructure.

Vanishing Hawking temperature enables one to focus on a crucial first step: the ``information storage problem.''  Supersymmetry allows one count states using index methods when Newton's constant vanishes. This appears radical, but index states are protected by supersymmetry under smooth deformations.  Therefore the index states will be preserved as one turns on $G_N$. Thus, even if one does not know what the microstructure becomes in the black hole, one knows that it must still be there, somewhere.  

However, to resolve the information paradox, it is not sufficient to count the microstates; one must determine their {\it structure} at finite  $G_N$, where the classical black-hole solution with a large horizon area exists. The challenge here is that gravitational attraction is universal and gets stronger with $G_N$, and so matter gets ever more compressed as $G_N$ increases.  On the other hand, the area of a black-hole horizon grows with $G_N$.  Thus, one would na\"ively expect that the index states will collapse behind a horizon \cite{Horowitz:1996nw,Damour:1999aw} as gravitational compressive forces exceed the ability of matter to sustain the pressure necessary to prevent core collapse. 

String theory evades this seemingly inevitable outcome through {\it fractionation} of its spectrum on extended objects \cite{Das:1996ug,Maldacena:1996ds}:

String theory necessarily contains solitonic objects: branes.  When a brane, or string, is bound to a large number, $N$, of another set of branes, then the tension of the first brane becomes a fraction, $1/N$, of its normal tension.  In  \cite{Mathur:1997wb} an estimate was made of the size of the brane bound states of Strominger and Vafa.  Remarkably, it was found that the size $D$ of this bound state increased with the number of branes in the state as well as with the coupling, such that we always have $D\sim R_s$, where $R_s$ is the radius of the horizon. 
Similar results were found~\cite{Banks:1997hz,Horowitz:1997fr,Banks:1997tn} in the BFSS matrix model~\cite{Banks:1996vh}, a precursor to (and in hindsight, an example of) gauge/gravity duality.
These results suggested that in string theory one never gets the traditional black hole with a vacuum horizon; instead one gets a horizon-sized `fuzzball,' which radiates from its surface like a normal body, thus evading the information paradox.  Indeed, it was subsequently shown  \cite{Bena:2004wt} that the gravitational back-reaction implies that the types of stringy configurations counted in \cite{Strominger:1996sh} grow with $g_s$ at exactly the same rate as the horizon.  Thus, at finite $g_s$, stringy microstructure does not necessarily form horizons.

Shortly after this came the AdS/CFT correspondence and holographic field theory, which also suggested that black-hole formation and evaporation must be a unitary process.  Essentially, the formation of a black hole (at least in AdS) is dual to some evolution of states in a holographic theory on the boundary of the space-time.  Since the dual field theory is unitary, the gravitational process must also be unitary.  

Thus, given the AdS/CFT correspondence, one has a wide variety of examples of unitary quantum dynamics of black holes.  As such, this correspondence provides an {\it existence} proof of counterexamples to Hawking's assertion that black holes violate quantum mechanical unitarity.  However, this result is not wholly satisfactory; what one really wants is a {\it constructive} proof, one that tells us {\it how the bulk gravitational dynamics is modified so that it becomes unitary}.  As one solves this problem, one can also try to match the unitary evolution in the bulk with that of  the boundary, and thus develop the holographic dictionary.

Fuzzballs and microstate geometries fit in the AdS/CFT duality in an obvious and natural way. The CFT has states that we can identify as states of a `black hole' phase. The gravity dual of any such microstates must be a solution of string theory with no horizon.  The absence of a horizon is essential for any individual microstate; horizon area is related to entropy in black hole thermodynamics, and an individual microstate has no entropy \cite{Lunin:2001jy,Lunin:2002qf,Mathur:2014nja}.  

This picture was also strongly supported by developments in holographic field theory. Black holes, with their horizons and temperature, provide the gravity duals of thermal field theory and ensemble averaging. On the other hand, individual phases or states, when sufficiently coherent, are described by smooth geometries.  For example, the confining phase of a theory that is in the same universality class as QCD is dual to the smooth Klebanov-Strassler geometry \cite{Klebanov:2000hb}. More generally, one of the core ideas of string compactification re-surfaced in holography~-- the link between phase changes in the field theory and geometric transitions.  A singular brane source, representing one possible phase, can ``dissolve'' into  fluxes that thread new topological cycles that emerge in the core of the transitioned geometry. The fluxes on these cycles are holographically dual to the order parameters of  the emergent infra-red phase of the field theory, and the scale of the cycles provides the  scale of the new phase \cite{Polchinski:2000uf, Bena:2004jw, Lin:2004nb, Dymarsky:2005xt}.  Thus gaugino condensates and chiral symmetry breaking are described by precisely such a transition from a singular brane-sourced geometry to a smooth, topologically non-trivial background \cite{Klebanov:2000hb}.

Microstate geometries have their roots  \cite{Giusto:2004kj,Bena:2005va,Berglund:2005vb,Bena:2006is,Bena:2007kg,Bena:2013dka} in these holographic insights.  The core idea is that, whatever the the black-hole microstates become at finite $G_N$, ground states and sufficiently coherent excitations can often  be described by smooth solutions to low-energy limits of string theory: supergravity.  

\subsection{Fuzzballs and microstate geometries}
\label{ss:FMG}

It is important to distinguish the following two notions:
\begin{itemize}
\item
{\it Microstate geometries} are smooth horizonless solutions of the supergravity theory that is the effective long-wavelength approximation to string theory.  These solutions have the same mass, charge and angular momentum as the corresponding black hole in GR.
\item
{\it Fuzzballs} are the general class of objects in which the black-hole interior is supplanted by some new, quantum, horizon-scale structure having itself no horizon.  This structure is built out of the branes and other stringy ingredients that are known to account for the black-hole entropy at weak string coupling. 
\end{itemize}
In particular, microstate geometries are examples of fuzzballs, but not all fuzzballs need be microstate geometries.  The ``fuzz'' of fuzzballs might, for instance, include string or brane condensates whose wave-function extends over the horizon scale, and whose degrees of freedom are the dominant components of black-hole entropy.  The ultimate role of microstate geometries will boil down to how generic such geometries are in the ensemble of all microstates and whether there might be effects beyond supergravity that are crucial ingredients of horizon-scale physics. 

 It is now understood that black holes are highly chaotic systems~\cite{Shenker:2013pqa,Maldacena:2015waa}, and this is consistent with idea of a fuzzball as a horizonless, complex, stringy quantum state.  However,  such chaos is in tension with the idea of microstate geometries, which are smooth coherent examples of fuzzballs.  The practical point is that generic, chaotic quantum states are very hard to compute, and microstate geometries, and their excitations, provide an invaluable starting point for computing properties of horizon-scale microstructure.


\newpage 

The {\it central precept of fuzzballs} or {\it ``fuzzball principle:"}  is: 

\nicebox{Fuzzballs represent a new phase  that emerges when matter is compressed to black-hole  densities, and this new phase prevents the formation of a horizon or singularity.  A fuzzball does not have an information problem  because the internal states of the fuzzball are in causal communication with distant observers.  Horizons and singularities only appear if one tries to describe gravity using a theory that has too few degrees of freedom to resolve the physics.}

\noindent
Hence, the idea is that string theory has sufficiently many degrees of freedom to resolve individual black-hole microstates into complicated, intrinsically-stringy, horizonless  objects: fuzzballs.  While vastly more complicated, fuzzballs have no more of an information problem than do pieces of coal, white dwarfs or neutron stars.

Microstate geometries are then the coherent states within the ensemble of all black-hole, or fuzzball, microstates.    While obviously more limited than generic fuzzballs, microstate geometries have the huge advantage that detailed computations can be done, and they can give invaluable insights into more generic fuzzballs.

\subsubsection{Fuzzball formation}
\label{ss:FuzzFormation}

One of the most natural challenges to the whole idea of fuzzballs is to consider a thought experiment whose outcome appears to be determined by some generalization of Birkhoff's theorem.  Specifically, arrange the collapse of a vast, perfectly-spherical shell of low-density matter, but with a titanic total mass.  How could this collapse into anything other than a Schwarzschild black hole? How  could perfect spherical symmetry not inevitably lead  to a horizon and a singularity?  More generally, General Relativity suggests that there are a vast array of initial states that must collapse to form a horizon and singularity.

In a similar vein, we also  learn in General Relativity that the horizon region is in the vacuum state, and that the curvature at the horizon of a black hole of mass $m$ is proportional to $m^{-2}$.  Thus, for large black holes, the space-time in the horizon region is very close to being empty, flat space, and must therefore be completely normal.  

We believe that both of these classical intuitions are too na\"ive because they ignore the elephant in the room~-- specifically, the  huge entropy, and the concomitant vast concentration of low-energy excitations sitting in the black hole.  

  The Bekenstein entropy \cite{Bekenstein:1973ur}
  \be
S_{\it Bek}={A\over 4G}
\label{SBek}
\ee
   is {\it much} larger than the entropy of normal matter that we could place within the same region with the given energy. But with the traditional picture of the black hole, it is not clear what role this large entropy could play in the dynamics of the black hole. With fuzzballs, there is a small but non-zero probability ${\mathcal P}\sim e^{-S_{\it Bek}}$ for the collapsing star to tunnel into a fuzzball configuration, which is compensated by the abnormally large density of states ${\mathcal N}\sim  e^{S_{\it Bek}}$~\cite{Mathur:2008kg}.  In \cite{Kraus:2015zda} an indirect argument was given for the exact cancellation of exponents in  ${\mathcal P}$ and ${\mathcal N}$. In \cite{Bena:2015dpt} an explicit example was worked out for this `entropy enhanced tunneling' within a certain family of fuzzball states.
  
Thus the answer to the issue of fuzzball formation is that, during collapse, a fuzzball must emerge through spreading of the quantum wave-function, governed by a quantum phase transition\footnote{Note that this entropy-enhanced tunneling is unique to fuzzballs, which are expected to have the same entropy as the black hole, and this distinguishes them from all other proposed black-hole-sized horizonless Exotic Compact Objects (ECO's).  Since such objects have vanishingly small entropy, they are not quantum incompressible in the way that the ensemble of fuzzballs is.}.  The vast density of microstates means that a black hole becomes a quantum object at the horizon scale: 

\nicebox{When one compresses matter into a state in which its entropy approaches $\frac{1}{4} A$, where $A$ is its surface area in Planck units, then it becomes an intrinsically quantum object. Thus, quantum effects become large at the horizon scale.}

\noindent
Such {\it quantum incompressibility} is one of the most important aspects of  black-hole physics.   It also implies that   the space-time, at the horizon scale, is far from classical ``normality.''

\section{Developments in fuzzballs and microstate geometries }
\label{Sect:Developments}

On one level, the  fuzzball paradigm is very conservative: it resolves the information problem with non-trivial structure at the horizon scale.  However, to achieve this within string theory requires an ambitious, even radical, re-thinking of  black-hole physics. Despite the challenges, the fuzzball paradigm rests on over two decades of sustained achievement, with results that have opened up new areas of investigation, provided new tests of the program and continue to give new insights into the fundamentals of black-hole microstructure.

In this section we will summarize some of the major milestones of this research and how it shaped the vision expounded in the previous section.

\subsection{``Microscopic'' microstate geometries}
\label{ss:micromicro}

The construction of fuzzballs started with perhaps the simplest black hole, one that is supersymmetric and ``microscopic'' in that it had a string-scale horizon area \cite{Sen:1995in}.  {\it All} its microstates could be explicitly constructed as limits of supergravity solutions \cite{Lunin:2001jy,Lunin:2002iz,Taylor:2005db,Kanitscheider:2007wq}.  Quantizing the space of these solutions yielded the Bekenstein entropy \cite{Rychkov:2005ji}, and {\it no microstate had a horizon}.   The  surface area, $A$, of the region where the microstates differ from the black hole satisfies a Bekenstein relation, $S_{\it Bek}\sim A/G$.  

These solutions provided a precise illustration of how holography should apply to black holes: Each of the microstates  can be put into a one-to-one correspondence with a state of the dual CFT\cite{Lunin:2001jy}.  Thus black-hole microstates are, in principle, no different than any other state of the theory, and we can think of a black hole as just a `string star'.  This was the beginning of the fuzzball paradigm. Similar results were also obtained for black strings \cite{Lunin:2002qf}.

However, these first steps were achieved with microscopic microstate geometries.  To build a more compelling picture of black-hole microstructure within the fuzzball paradigm, one must first construct ``macroscopic'' microstate geometries.

\subsection{``Macroscopic'' microstate geometries}
\label{ss:macromicro}

``Macroscopic'' microstate geometries are those that correspond to black holes whose horizon area is parametrically large.  At first sight, it appeared that finding such geometries would be impossible because they would involve all the non-linear interactions of supergravity.  Moreover there were  an array of ``no-go'' theorems that suggested that such macroscopic microstate geometries could not exist.

However, a series of breakthroughs \cite{Bena:2004de,Bena:2005va,Berglund:2005vb,Bena:2011uw,Bena:2011dd,Shigemori:2013lta,Ceplak:2018pws,Heidmann:2019zws,Mayerson:2020tcl} not only made the solution of this problem possible, but enabled the systematic construction of vast families of such microstate geometries\cite{Bena:2006kb,Bena:2007qc,Bena:2007kg,Bena:2008wt,Bena:2010gg,Bena:2015bea,Bianchi:2016bgx,Bena:2016ypk,Bianchi:2017bxl,Heidmann:2017cxt,Bena:2017geu,Avila:2017pwi,Bena:2017xbt,Tyukov:2018ypq,Heidmann:2019xrd,Warner:2019jll,Shigemori:2021pir,Ganchev:2021pgs,Ganchev:2021ewa}.  Dissecting the ``no-go'' theorems revealed that the mechanism underlying microstate geometries is the {\it only} way to create non-trivial, horizonless, time-independent solitonic solutions with black-hole mass and charges \cite{Gibbons:2013tqa,deLange:2015gca}.

Most important were the ``scaling'' microstate geometries \cite{Bena:2006kb,Bena:2007qc,Denef:2007vg,deBoer:2008zn,deBoer:2009un,Vasilakis:2011ki,Martinec:2015pfa} for which there is a parameter that controls the red-shift at which the soliton deviates significantly from the black hole and ``caps-off smoothly"  at the horizon scale.  Semi-classical quantization of the moduli space of these geometries limits the ``depth'' of the black-hole-like throat, giving the microstate geometry a maximal red-shift and an energy gap \cite{Bena:2006kb,Bena:2007qc,deBoer:2008zn,deBoer:2009un}.  This energy gap matched exactly with that of the maximally fractionated sector of the dual holographic field theory \cite{Bena:2018bbd}, and showed, for the first time, that microstate geometries could access the sector of the holographic field theory that contributes most to the entropy of the corresponding black hole. 

This was the first step in what is now a well-developed program of {\it precision holography} that gives a precise dictionary between a large class of scaling microstate geometries and excited states of the dual holographic field theory that underlies the black-hole microstructure. 

\subsection{Precision holography}
\label{ss:prechol}

For the last decade there has been a remarkable symbiosis between microstate geometries and the holographic field theory of the D1-D5 brane system.  Advances in the holographic field theory  \cite{Kanitscheider:2006zf,Kanitscheider:2007wq,Taylor:2007hs,Giusto:2015dfa,Bombini:2017sge,Giusto:2019qig, Tormo:2019yus, Rawash:2021pik} pointed to  essential geometric degrees of freedom needed to describe the gravitational realization of some of the index states of Strominger and Vafa, and led to the breakthrough construction of geometries  known as {\it superstrata}~\cite{%
Bena:2015bea,
deLange:2015gca,
Bena:2016agb,
Bena:2016ypk,
Bena:2017xbt,
Bakhshaei:2018vux,
Ceplak:2018pws,
Heidmann:2019zws,
Shigemori:2019orj,
Heidmann:2019xrd,
Mayerson:2020tcl,
Shigemori:2020yuo,
Mayerson:2020acj,
Ganchev:2021iwy}.   
Conversely, breakthroughs in the construction of new families of superstrata have led to predictions for correlation functions in the holographic field theory.  Microstate geometries have passed all these tests~\cite{%
Giusto:2015dfa,
Galliani:2016cai,
Bombini:2017sge,
Bombini:2019vnc,
Tian:2019ash,
Tormo:2019yus,
Giusto:2019qig,
Bena:2019azk,
Giusto:2019pxc,
Hulik:2019pwr,
Giusto:2020mup,
Ceplak:2021wzz,
Rawash:2021pik}, 
and there is now a huge body of evidence that the superstrata are indeed capturing black-hole microstructure, and, in particular, some of the index states.

\subsection{Black-hole-like behavior}
\label{ss:Behavior}

For over a decade, we have had microstate geometries that look like black holes except very close to the horizon scale, at very high red-shift, where microstate geometries cap-off smoothly. For the last five years there have been many probes of, and computations within, these geometries to reveal the similarities with, and differences from, black holes.  Some highlights of this work include the following:
\begin{enumerate}[\it (i)]
\setlength\itemsep{-0.1em}
\item Probing microstate geometries with waves \cite{Eperon:2016cdd,Marolf:2016nwu, Bena:2017upb, Bianchi:2017sds, Bena:2018bbd, Raju:2018xue,Bianchi:2018kzy,Bena:2019azk,Chakrabarty:2019ujg, Bena:2020yii, Craps:2020ahu,Bianchi:2020des,Bianchi:2020yzr,Bacchini:2021fig,Ikeda:2021uvc,Craps:2021bmz,Bianchi:2021xpr, Bianchi:2021mft} reveals  black-hole-like decay of the wave over the short term, as the wave ``falls into'' the geometry. After long periods of time the wave ``echoes'' off the bottom of the geometry, and the ``information`` in the wave is returned. 

\item Microstate geometries have a rich collection of bound states that localize at the bottom of the deep throats.  However, these states can tunnel to flat space far from the core of the geometry.  The amplitude for tunneling is extremely small, and it is believed that this could yield the analog of Hawking radiation for microstate geometries   \cite{Chowdhury:2007jx,Chakrabarty:2015foa,Chakrabarty:2019ujg,Bena:2020yii}.

\item The ultra-relativistic infall speeds  of infalling particles, combined with the tiny deviations of microstate geometries from black holes, causes a probe to experience string-scale tidal forces \cite{Tyukov:2017uig, Bena:2018mpb, Hampton:2019csz,Bena:2020iyw,Martinec:2020cml,Ceplak:2021kgl,Guo:2021ybz,Guo:2021gqd}.  Indeed, a similar process was already evident in ultra-relativistic string scattering \cite{Amati:1987wq,Amati:1987uf,Giddings:2006vu,DAppollonio:2010ae}. The tidal forces are sufficient to scramble a string probe (including the massless strings of (i)) into its excited  states, converting kinetic energy into mass, thus trapping the particle deep within the geometry \cite{Martinec:2020cml,Ceplak:2021kgl}.   This {\it tidal trapping} is the first step in the black-hole-like scrambling of matter within microstate geometries.  This tidal trapping also dramatically diminishes the return``echoes'' of (i). 

\item  Because microstate geometries have structure, their   gravitational  multipole moments  differ from those of black holes.  These multipoles have been extensively characterized  and quantified \cite{Bena:2020see,Bianchi:2020bxa,Bena:2020uup, Bianchi:2020miz, Bah:2021jno, Fransen:2022jtw} in the hope that they might ultimately provide some observational signature of microstate structure \cite{Mayerson:2020tpn,Mayerson:2022yoc}.
\end{enumerate}
%

\subsection{Strings and fuzzballs}
\label{ss:strings}

The passage from microstate geometries to generic fuzzballs requires string theory rather than its massless limit, supergravity.  The first step is to study perturbative string theory effects in the simplest non-trivial microstates, where the supergravity approximation breaks down in the ``cap'' of the geometry because of string-scale curvature.  
The world-sheet formalism yields a perturbative expansion about a given microstate background, and can in principle handle such highly curved backgrounds.   

Fortunately, many of the particular families of microstates most studied in the literature~\cite{%
Lunin:2001fv,
Giusto:2004id,
Jejjala:2005yu,
Giusto:2012yz,
Chakrabarty:2015foa} 
admit an exactly solvable world-sheet dynamics in which the background is sourced by NS-NS three-form fields, fundamental strings and NS5-branes%
~\cite{Martinec:2017ztd,Martinec:2018nco,Martinec:2019wzw,Martinec:2020gkv,Bufalini:2021ndn}.  
The analysis provides a picture of what happens when these particular backgrounds are deformed to more generic configurations.

String dynamics plays an essential role in resolving the singular regions of the supergravity moduli space of NS5-branes.  When the latter are slightly separated by sub-string scale distances, these singularities are resolved within perturbative string theory~\cite{Giveon:1999px,Giveon:1999tq}; however coincident NS5-branes generate a strong-coupling singularity that is only resolved by the appearance of light stretched or wrapped D-branes, as has been seen in related contexts~\cite{Hull:1994ys,Strominger:1995cz,Witten:1995zh}.  

These phenomena play out in the singular limits of the backgrounds of  Section~\ref{ss:micromicro}, where they appear in the deeply red-shifted cap of a microstate geometry sourced by a much more complicated and generic five-brane configuration~\cite{Martinec:2019wzw,Martinec:2020gkv}.  Over most of the space of microstates, the five-branes are separated, and the stringy structure of the cap is resolved perturbatively.  Where that breaks down, the new light branes that arise are non-abelian strings bound to the five-branes, revealing {\it in the bulk dynamics} the sorts of degrees of freedom counted in~\cite{Strominger:1996sh,Dijkgraaf:1996cv,Maldacena:1996ya}, and providing an explicit example of the sorts of fractionation effects that are key to understanding black-hole entropy and generic fuzzball states.

\subsection{Non-BPS microstate geometries}
\label{ss:nonBPS}

Supersymmetry, and the associated first-order BPS equations, greatly simplify the construction of microstate geometries, and so by far the most progress has been made in  supersymmetric constructions.  Moreover, precision holography has focused  on supersymmetry because it protects a range of correlators that can be computed on both sides of the duality.   However, astrophysical black holes are far from BPS  and so it is essential for the microstate geometry program to make the leap to non-BPS geometries.

Over the years, there has been some construction of non-supersymmetric microstate geometries, but these have often led to extremal solutions, or solutions with very high angular momenta  \cite{Jejjala:2005yu,Bena:2009qv,Bobev:2009kn,Bobev:2011kk, Vasilakis:2011ki, Bena:2015drs, Bena:2016dbw, Bossard:2017vii, Heidmann:2018mtx, Bah:2020ogh, Bah:2020pdz, Bah:2021owp}.  However, the last year has seen some spectacular progress in the systematic construction of very interesting, more physical classes of non-supersymmetric microstate geometries \cite{Ganchev:2021pgs,Ganchev:2021ewa, Bah:2021rki, Heidmann:2021cms, Bah:2021irr}.  These constructions are quite new, but they underline how some of the remarkable features of supersymmetric microstate geometries are shared by non-supersymmetric counterparts.   The holography of some of these geometries is already being developed \cite{Ganchev:2021ewa}, and extending these constructions is already underway.  The spectrum of  excitations around these geometries is  revealing how a generic non-BPS microstate geometry might make the transition to chaos.    Above all, this new work has opened up a vast new area for future research.

\section{Outlook }
\label{Sect:Outlook}

The fuzzball  and  microstate geometry programs have been growing and developing for over two decades and
have made substantial progress in understanding the horizon structure of black holes and resolving the black-hole information paradox.  

The foundation of this work lies in a rich circle of ideas within string theory, and is supported by detailed, and sometimes challenging, computations.  Each new result  suggests further computations to test, or build out, the overall picture of black-hole microstructure, adding to the forward momentum of the research.  One of the core strengths of the  fuzzball and microstate geometry programs is that they lie at the intersection of many different areas of fundamental physics: field theory, string theory, supergravity, GR and quantum information theory.  Each of these perspectives informs the other and many advances have been propelled by such symbioses.

As we have tried to describe in this white paper, the fuzzball  and  microstate geometry programs provide one of the best approaches to understanding the physics black-hole microstructure.  Based on the current level of activity, and the progress to date, these programs will remain a vibrant and exciting area of future research.  There remain many open questions and avenues to explore, especially in taking what we have learned so far and using it to inform our knowledge of astrophysical black holes.

\section*{Acknowledgments}
\vspace{-2mm}
We would like to thank our many collaborators and colleagues whose work and insights enabled us to see a little further.  
The work of NW is supported in part by the DOE grant DE-SC0011687. 
The work of EJM is supported in part by DOE grant DE-SC0009924.
The work of  IB and NW is supported in part by the ERC Grants 787320 - QBH Structure and 772408 - Stringlandscape.
The work of  SDM  is supported in part  by DOE grant DE-SC0011726.


\begin{adjustwidth}{-1mm}{-1mm} 

\bibliographystyle{utphys}

\bibliography{Snowmass-Fuzz-final}       

\end{adjustwidth}

\end{document}